\documentclass{aa}  

\usepackage{graphicx}
\usepackage{txfonts}
\usepackage{soul}
\usepackage{color}

\begin{document}

   \title{Photospheric magnetic topology of a north polar region}

   \author{A. Pastor Yabar
          \inst{1}
          \and
          M. J. Mart\'{\i}nez Gonz\'alez
          \inst{2,3}
          \and
          M. Collados
          \inst{2,3}
          }

   \institute{Leibniz-Institut f\"ur Sonnenphysik (KIS),
             Sch\"oneckstr. 6, 79104 Freiburg, Germany
             \email{apy@leibniz-kis.de}
         \and
             Instituto de Astrof\'{\i}sica de Canarias,
             C. V\'{\i}a L\'actea s/n., 38205 La Laguna, Tenerife, Spain\\
         \and
             Departamento de Astrof\'{\i}sica de Canarias, Universidad de La Laguna,
             Avda Astrof\'{\i}sico S\'anchez s/n., 38205, La Laguna, Tenerife, Spain
             }

   \date{Received September 15, 1996; accepted March 16, 1997}

 
  \abstract
   {}
   {We aim to characterise the magnetism of a large fraction of the north polar region close to a maximum of activity, when the polar regions are reversing their dominant polarity.}
   {We make use of full spectropolarimetric data from the CRisp Imaging Spectro-Polarimeter installed at the Swedish Solar Telescope. The data consist of a photospheric spectral line, which is used to infer the various physical parameters of different quiet Sun regions by means of the solution of the radiative transfer equation. We focus our analysis on the properties found for the north polar region and their comparison to the same analysis applied to data taken at disc centre and low-latitude quiet Sun regions for reference. We also analyse the spatial distribution of magnetic structures throughout the north polar region.}
   {We find that the physical properties of the polar region (line-of-sight velocity, magnetic flux, magnetic inclination and magnetic azimuth) are compatible with those found for the quiet Sun at disc centre and are similar to the ones found at low latitudes close to the limb. Specifically, the polar region magnetism presents no specific features. The structures for which the transformation from a line-of-sight to a local reference frame was possible harbour large magnetic fluxes ($>10^{17}$ Mx) and are in polarity imbalance with a dominant positive polarity, the largest ones  ($>10^{19}$ Mx) being located below $73^{\circ}$ latitude.}
   {}

   \keywords{Sun: magnetic fields --
                Sun: photosphere --
                Techniques: polarimetric
               }

   \maketitle
%

\section{Introduction}

A particularly interesting feature of the Sun's magnetic field is its cyclic evolution. The solar activity cycle exhibits an average period of 11 years \citep{schwabe1844}. In each period, the Sun undergoes activity maxima with a large number of magnetic events (sunspots, flares, coronal mass ejections, etc.), and activity minima when magnetic structures are hardly noticeable and the solar photosphere is mostly quiet. If we take into account the polarity of the magnetic field, this period is doubled as two consecutive activity cycles present opposite polarity, that is, the polarity of the active regions (defining the polarity of the active region as that of the leading part) is reversed between two consecutive maxima \citep[known as the Hale law;][]{hale1925}. Moreover, the dominant polarity of each polar region changes from one minimum to the next \citep{babcock1959}. In this context, the  magnetism of the solar poles is usually employed as a proxy for the strength of the next solar cycle \citep[e.g. see section 2.2 in][and references therein]{petrovay2010} and polar regions are also relevant in the extrapolation of the magnetic field into the corona and the heliosphere as polar caps host most of the solar open magnetic fields \citep{petrie2015}. Therefore, measurements of the magnetic properties of the solar poles are crucial for understanding the Sun's magnetic cycle.

The solar poles are generally referred to as the regions above $60^\circ$ latitude, where there is no emergence of active regions and the surface is quiet. It is therefore expected that the polar magnetism shares some similarities with the small-scale, weak magnetism that populates the quiet Sun (QS). However, the lack of active regions and the close relation to the solar cycle leads us to postulate that some fundamental differences might exist between the magnetism of the polar caps and that of the more equatorial regions. Full characterisation of the polar magnetic vector became possible when polarimetric observations were made of the full Stokes vector in polar areas. These observations are extremely challenging mainly because of the foreshortening effect that makes it difficult to obtain the high spatial resolution required to disentangle the small-scale magnetism of specific regions close to the solar limb as observed from the Earth. Also, interpretation of the observed signals is difficult because the geometry of  observations close to the limb forces us to solve the fundamental ambiguity of the plane-of-the-sky azimuth of the magnetic field to infer the geometry of the field and, more importantly, to determine the polarity of the field. 

Despite these difficulties, several groups have attempted the study of the polar magnetism with the help of high-quality spectro-polarimetric observations. \cite{tsuneta2008}, using high-spatial-resolution data from the Hinode satellite, found that the polar magnetism was characterised by strong and vertical magnetic field patches scattered throughout the polar cap. The polarity of such vertical structures coincided with that of the global field. With similar data, \cite{ito2010} compared the magnetism of the polar region with the QS at the equatorial limb (i.e. projection effects are the same in both data sets). These latter authors found that the weaker magnetic fields at both disc positions had similar topologies but that the strongest vertical features had an imbalance of polarity at the polar caps that was not found at the equatorial limb. As in \cite{tsuneta2008}, the polarity of these strong fields was the same as the dominant polarity of the polar region. Similarly, but using ground-based data, \cite{blancorodriguez2010} found  strong and vertical fields at the poles harbouring the polarity of the global field, and weak, more isotropic and polarity-balanced magnetic fields. \cite{shiota2012} studied the evolution of the polar region magnetism over four years  and found that the number of large vertical patches (specific to the polar
regions according to \cite{ito2010}) varies with time. The closer to the activity minimum, the larger the number of such structures and, conversely, close to activity maximum, the number of those magnetic concentrations in the polar areas is the smallest. More recently, \cite{pastoryabar2018} also agreed that some magnetic fields at the polar regions are characterised by  strong vertical fields, but those authors also found some polarimetric signals common to polar regions and the equatorial limb that were compatible with the presence of unresolved small-scale magnetic loops.

Despite these important results appearing during the last decade, we still do not fully understand the magnetism of the polar regions. Moreover, although some studies have attempted to solve this problem \citep[][and references therein]{charbonneau2010}, the mechanism that provides the polarity reversal at the polar caps is still not characterised from an observational point of view. Here, we  study the physical properties (mainly focusing on the topology of the magnetic field) at the polar regions by means of high-spatial-resolution spectro-polarimetric observations close to a maximum of activity and covering a
large area. During this period, polar regions are close to reversing their polarity, a key point for understanding the magnetism of these regions.
%

\section{Observations and data reduction}

On August 19 2015 we recorded full vector spectropolarimetric data of the north polar region with the CRisp Imaging Spectro-Polarimeter \citep[CRISP,][]{scharmer2006,scharmer2008} installed at the Swedish Solar Telescope \citep[SST;][]{scharmer2003} at the {\it Observatorio del Roque de los Muchachos} (Spain). At that time, the Sun's rotation axis was tilted so that the north polar region was well inside the visible disc ($B_{0}=6.8^{\circ}$), allowing the best chance to study its physical properties. The observations consisted of sequential scans of the photospheric Fe{\sc I} 6173 {\AA} and the chromospheric Ca {\sc II} 8542 {\AA}. Here, we focus only on the analysis of the Fe{\sc I} line, which provides information on the photosphere. This spectral line is very sensitive to the magnetic field, having a Land\'e factor of 2.5. The Fe{\sc I} line was scanned at 20 wavelength positions: from -225 m{\AA} to +225 m{\AA} with steps of 25 m{\AA}, and an additional spectral point at the continuum at 525 m{\AA} from the line core. The 0 m{\AA} reference was set to the position of the minimum of the spectral line as measured during the etalon calibration process in the morning (UT 07:10). At each wavelength position four modulations of the light were taken in order to recover the four Stokes parameters. Each of these four measurements at each wavelength position was repeated 12 times. The integration time for each image was close to 18 ms, and together with the 17 ms reading time, this spectral line was scanned in $\sim$32 seconds. The scan of this photospheric line together with the scan of the Ca {\sc II} 8542 {\AA} spectral line, which took $\sim$16 seconds, gave a sequential scanning cadence of $\sim$50 seconds. At each disc position we repeated this scanning setup between five and ten times as detailed in col. 7 of Table \ref{Tab:data}.

\begin{figure*}
  \centering
  \includegraphics[width=0.95\textwidth]{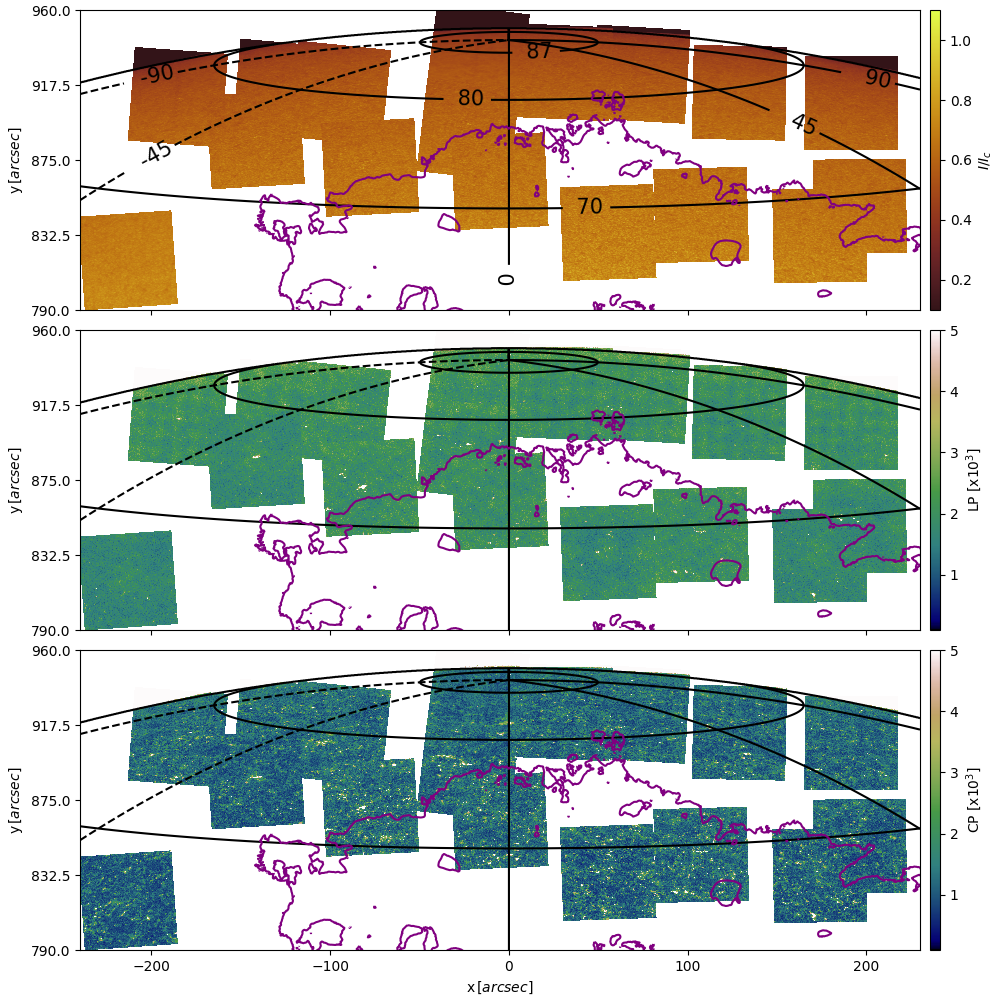}
  \caption{Mosaic of the north polar region with the 17 scans taken. The continuum intensity is shown in the upper panel, the linear polarisation map is shown in the middle panel, and the circular polarisation map is shown in the bottom panel. Linear polarisation is given by $\int_{\lambda}\sqrt{Q(\lambda)^2+U(\lambda)^2}\,d\lambda/\int_{\lambda}I(\lambda)\,d\lambda$ and circular polarisation is given by: $\int_{\lambda}|V(\lambda)|\,d\lambda/\int_{\lambda}I(\lambda)\,d\lambda$. The purple contour depicts the position of a coronal hole as estimated from the SDO/AIA 193 {\AA} intensitygram.}
  \label{Fig:npfovdata}
\end{figure*}

\begin{table*}
  \caption{Observational details of the scans employed.}
  \label{Tab:data}
  \centering
    \begin{tabular}{lccccccccc}
      \hline\hline
      Time (UT) & $x_{\odot}$ ($^{\prime\prime}$) & $y_{\odot}$ ($^{\prime\prime}$) & $\lambda_{\odot}$ $(^{\circ})$ & $l_{\odot}$ $(^{\circ})$ & $\mu_{\odot}$ $(^{\circ})$ & N & \%$_{LIN}$ & \%$_{MAG}$ & Target\\
      \hline\hline
07:19:01 & 0.60 & 0.70 & 6.86 & 0.04 & 1.00 & 10 & 0.01 & 6.45 & DC \\
\hline
07:47:05 & -915.40 & -0.70 & 1.77 & -74.66 & 0.27 & 10 & 0.04 & 2.85 & EL \\
\hline
07:54:59 & -883.10 & -7.60 & 2.05 & -68.51 & 0.37 & 10 & 0.02 & 3.95 & EL \\
\hline
08:04:09 & -14.30 & 935.80 & 86.92 & -16.28 & 0.17 & 10 & 0.01 & 1.55 & NP \\
\hline
08:12:26 & -21.30 & 892.10 & 76.74 & -5.61 & 0.34 & 10 & 0.17 & 4.85 & NP \\
\hline
08:20:27 & -94.00 & 912.30 & 79.95 & -34.55 & 0.26 & 10 & 0.07 & 4.25 & NP \\
\hline
08:29:31 & -124.90 & 913.90 & 79.66 & -47.13 & 0.24 & 10 & 0.06 & 3.55 & NP \\
\hline
08:38:38 & -184.10 & 908.70 & 77.36 & -62.39 & 0.22 & 10 & 0.07 & 3.35 & NP \\
\hline
08:47:24 & 72.80 & 926.30 & 83.23 & 40.56 & 0.21 & 10 & 0.04 & 2.95 & NP \\
\hline
08:56:11 & 32.60 & 926.30 & 83.90 & 18.85 & 0.22 & 10 & 0.02 & 3.25 & NP \\
\hline
09:09:31 & 131.90 & 918.20 & 80.21 & 54.79 & 0.21 & 10 & 0.05 & 3.95 & NP \\
\hline
09:17:34 & 193.50 & 914.30 & 77.70 & 73.00 & 0.18 & 10 & 0.06 & 3.25 & NP \\
\hline
09:26:01 & 198.00 & 855.70 & 70.02 & 37.60 & 0.38 & 5 & 0.17 & 5.25 & NP \\
\hline
09:30:51 & 175.90 & 838.30 & 68.08 & 29.75 & 0.43 & 5 & 0.26 & 7.75 & NP \\
\hline
09:35:13 & 109.50 & 848.50 & 69.83 & 19.53 & 0.43 & 5 & 0.34 & 8.05 & NP \\
\hline
09:39:35 & 57.30 & 837.60 & 68.63 & 9.53 & 0.47 & 5 & 0.16 & 8.65 & NP \\
\hline
09:44:02 & -3.00 & 865.70 & 72.54 & -0.60 & 0.41 & 5 & 0.16 & 8.05 & NP \\
\hline
09:48:24 & -76.80 & 870.80 & 73.10 & -16.16 & 0.39 & 5 & 0.44 & 9.65 & NP \\
\hline
09:52:48 & -141.30 & 885.20 & 74.73 & -34.40 & 0.33 & 5 & 0.16 & 3.25 & NP \\
\hline
09:57:44 & -212.30 & 815.10 & 65.10 & -32.07 & 0.46 & 5 & 0.09 & 4.75 & NP \\
\hline
10:21:16 & 3.30 & -0.20 & 6.80 & 0.20 & 1.00 & 10 & 0.01 & 7.85 & DC \\
\hline
    \end{tabular}
\tablefoot{x ($x_{\odot}$) and y ($y_{\odot}$) helioprojected values of the central point of each FOV, the solar latitude ($\lambda_{\odot}$) and longitude ($l_{\odot}$) and $\mu=\cos\theta$, where $\theta$ is the heliocentric angle. N is the number of repetitions of the scans of both lines at each FOV. \%$_{LIN}$ is the percentage of the FOV that presents linear polarisation signatures above the noise level for the Fe{\sc I} line. \%$_{MAG}$ is the percentage that shows any, linear, or circular polarisation above the signal criterion for that spectral line. {\it Target} is a label to easily identify the FOV of the scan: NP for north pole, EL for east limb, and DC for disc centre.}
\end{table*}

\begin{figure*}
  \centering
  \includegraphics[width=0.85\textwidth]{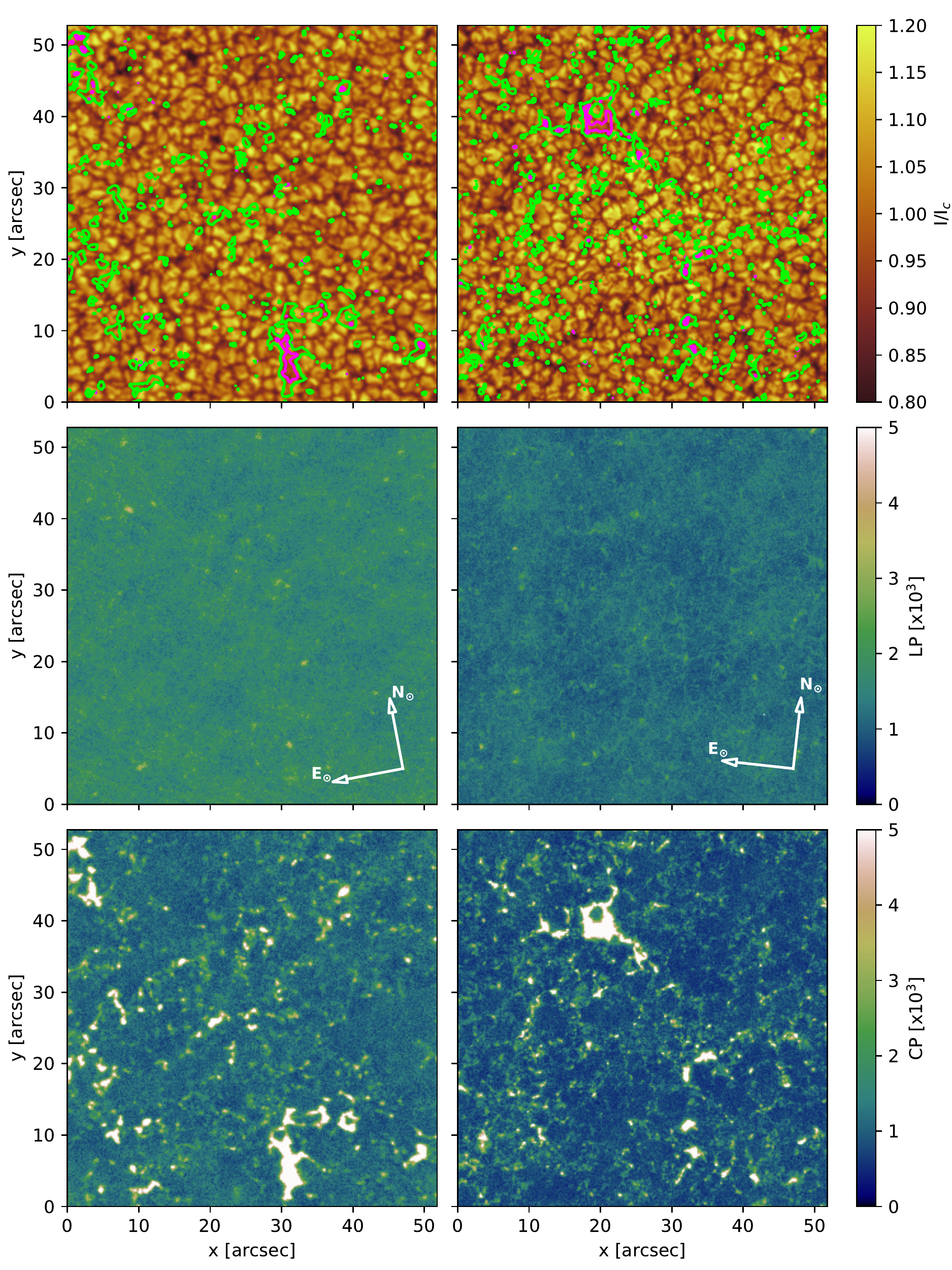}
  \caption{Disc centre observations. From top to bottom: Intensity, linear polarisation, and circular polarisation maps taken at UT 07:19:01 (left) and at UT 10:21:16 (right). Contours in intensity maps highlight the following subset of data: all the pixels that were inverted (green), pixels for which orientation is considered to be well determined (magenta).}
  \label{Fig:dcfovdata}
\end{figure*}

\begin{figure*}
  \centering
   \includegraphics[width=0.85\textwidth]{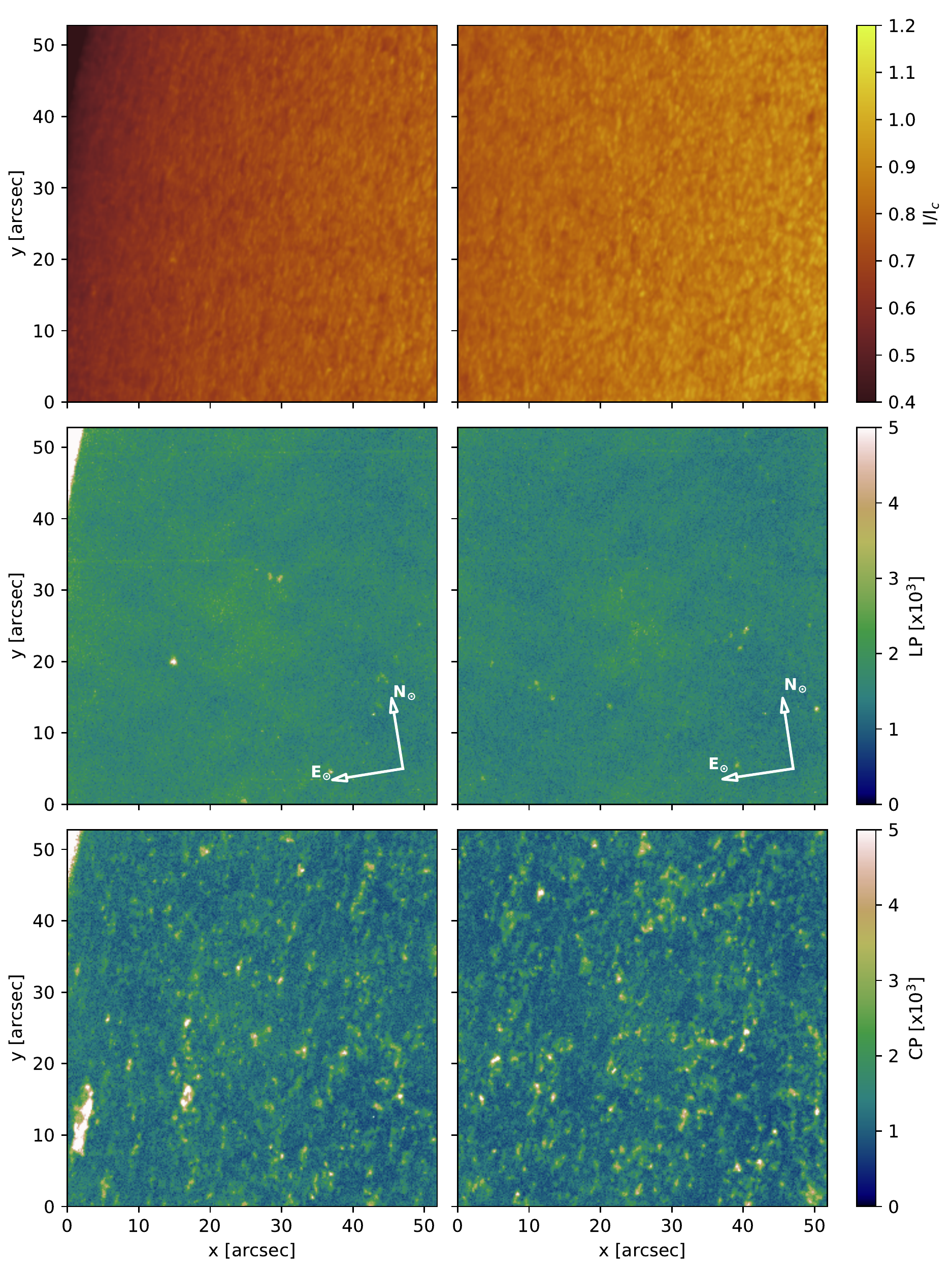}
  \caption{As in Fig. \ref{Fig:dcfovdata} but for two scans taken at the east limb: at UT 07:47:05 (left) and at UT 07:54:59 (right).}
  \label{Fig:elfovdata}
\end{figure*}

The excellent seeing conditions together with the use of the adaptive optics system \citep[which is an update of the previous one; ][]{scharmer2003}, allowed us to obtain 17 high-spatial-resolution maps (0.$^{\!\!\prime\prime}$35, see next paragraph) of a $\sim 50^{\prime\prime}\times 50^{\prime\prime}$ area that provided very good coverage of the north polar region (see Fig. \ref{Fig:npfovdata}). To compare the analysis of the polar region, we also took two $50^{\prime\prime}\times 50^{\prime\prime}$ maps of the QS at disc centre (Fig. \ref{Fig:dcfovdata}) and another one at the east limb (Fig. \ref{Fig:elfovdata}). In those two cases, the spatial resolution was as good as in the polar region.

The data reduction was performed using the dedicated CRISPRED pipeline \citep{delacruzrodriguez2015}. This software performs the dark current subtraction, flat-field correction, demodulation, the removal of residual cross-talk, the removal of polarimetric fringes, and the correction of the filter transmission profile. The standard reduction typically  also involves image restoration by means of Multi-Object Multi-Frame Blind Deconvolution \citep[MOMFBD; ][]{vannoort2005}. However, instead, we preferred to use a de-stretching module (kindly implemented by de la Cruz Rodr\' iguez), because the polarisation signals at QS regions are very weak and image restoration techniques tend to increase the noise level. This way, the reduced data did not reach the diffraction limit of the telescope (0.$^{\!\!\prime\prime}$12) but they achieved a spatial resolution of 0.$^{\!\!\prime\prime}$35, as retrieved from the two-dimensional Fourier power spectrum. Since the pixel size of our data was 0.$^{\!\!\prime\prime}059$, the maps were spatially oversampled. We then performed a binning of 3x3 pixels, ending with a pixel size of 0.$^{\!\!\prime\prime}177$ and an increase of the signal-to-noise ratio by a factor of three. Also, as we do not remove the effect of seeing and perform a binning over the data, we expect that these seeing and binning effects might be relevant in the inference of physical parameters. This is because these two effects mix information coming from various pixels, which typically leads to smoother spatial physical distributions. A straightforward parameter that provides insight into the consequences of these effects is the continuum contrast. For example, \cite{riethmueller2014}  analysed the effect of the  point spread function (PSF) of the telescope and the spectral PSF over a QS simulated region. These latter authors found that the simulated QS intensity contrast is reduced by the action of the telescope PSF and the instrumental PSF from 22\% to 12\% in the case of 5250 {\AA}. Here, as we have to additionally include the effect of seeing and binning, we expect that the measured intensity contrast is further reduced. In particular, we find that our QS disc centre scans taken at UT 07:19 and UT 10:21 have intensity contrasts of 6.65 \% and 7.07 \%, respectively (we note that a direct comparison of the values is not possible due to the different wavelength). These intensity contrast values can  also be  compared to the ones shown by \cite{scharmer2019} in which, among others, they compare the RMS contrast of QS images for several wavelengths as a function of the Fried $r_{0}$ parameter (see Fig. 6 in that work). These latter authors find that, for a similar wavelength (630 nm in contrast to our wavelength range 617.3nm) and for a similar image type (blue dots on the plot which are compensated for the diffraction limited PSF, while our data are not), an intensity contrast of 7 \% corresponds to an $r_{0}$ of about 8 cm at 500 nm, similar to our subfield averaged $r_{0}$ values at the same wavelength for disc centre scans: $~\sim8$ cm and $~\sim7$ cm.

Finally, we set a common continuum intensity reference for the various fields of view (FOVs) and corrected for intensity variation due to the Sun's elevation during observations. To do so, we compared the centre-to-limb variation (CLV) of the observations with a synthetic CLV of the 6173 {\AA} continuum. This allowed the estimation of a time-dependent factor required to set all the observed FOVs to the same intensity reference. In order to avoid sharp variations between consecutive scans, the time varying factor was fitted to a third-order polynomial and then each FOV was corrected accordingly. After this correction was applied, the intensity reference was set to that of the average QS at disc centre continuum.

After the reduction process, the polarimetric sensitivity was $\sigma=1.\times10^{-3}\,I_{c}$, as derived from the standard deviation over time of the Stokes Q, U, and V parameters at the continuum point. The symbol $I_c$ denotes the continuum intensity. The spatial binning increased this sensitivity to $\sigma=4.\times10^{-4}\,I_{c}$. In order to infer a reliable magnetic field, we restricted our study to those signals whose amplitude in any of the polarised Stokes parameters exceeded (at least one wavelength point) $3\sigma$. This step left us with 5.17 \% of the observed areas fulfilling this selection criterion, and only 0.12 \% of the FOVs had linear polarisation signatures above the noise level (see cols. 8 and 9 in Table \ref{Tab:data} for the percentages of each FOV).

\section{Analysis of the data}
\label{Sec:inversion}

We infer the physical properties of the solar atmosphere using inversion techniques applied to the spectropolarimetric data, focusing on the
magnetic field and line-of-sight (LOS) velocity. We use the inversion code Stokes Inversion based on Response functions \citep[SIR;][]{ruizcobo1992}. This code solves the radiative transfer equation (RTE) under the assumption of local thermodynamic equilibrium (LTE), an approximation that is accurate enough for most of the photospheric spectral lines, in particular for the one used here.

For the inversion strategy, we consider each resolution element as the combination of two atmospheres. The first one is magnetic and its free parameters are the temperature (with up to five nodes), and height independent LOS velocity, microturbulent velocity, magnetic field strength, inclination, and azimuth. The second component is \mbox{non-magnetised} and its free parameters are the LOS velocity and the microturbulent velocity. The temperature is fixed to be the same for both magnetic and \mbox{non-magnetic} atmospheres. The filling factor, which gives the relative weight of each of the previous components, is also set as a free parameter. This two-component model is accompanied by the theoretical CRISP spectral transmission profile calculated with dedicated software given by the instrument team. 
This profile also carries the information of the instrumental induced wavelength shift by means of the so-called {\it cavity maps}, a byproduct of the CRISPRED reduction process. The two-model-atmospheres strategy is necessary to reproduce the presence of differential LOS velocities between the Stokes I and Stokes V profiles. In addition, the different microturbulent velocities allow us to fit the width of the Stokes I profile and the width of the Stokes Q, U, and V profiles simultaneously.

This two-component model cannot reproduce the presence of asymmetric Stokes Q, U, and V profiles. These features are indicative of gradients of velocity and/or magnetic field along the LOS. In the ideal case of a constant magnetic field with a constant LOS velocity, Stokes Q and U profiles are symmetric and Stokes V profiles are antisymmetric. We find that this is not the case for our data; however, an LOS inversion model with a constant LOS velocity and a constant magnetic field is able to retrieve an average estimation of this LOS variation \citep{westendorpplaza1998}. Another case to take care of is the presence of such strong gradients that one of the lobes of the Stokes V profile is suppressed. Single-lobed Stokes V profiles are defined when only one of the lobes of the Stokes V profile is above the $3\sigma$ threshold. This happens to $\sim$1 \% of the pixels with polarimetric signals. In this situation, the inversion strategy proposed here leads to inaccurate results, and so single-lobed Stokes V profiles are discarded in the forthcoming analysis. The subset of pixels inverted for the disc centre case are highlighted in green in Fig. \ref{Fig:dcfovdata}.

Each pixel is inverted 50 times with initial random values of LOS velocity, microturbulent velocity, and magnetic field strength, inclination, and azimuth. The random values for the LOS velocities are in between \mbox{$\pm$ 2 km/s} and that of microturbulent velocities are between 0 km/s and 2 km/s. The initial guess temperature stratification is that of the VALC semi-empirical model atmosphere \citep{vernazza1981}. Initial magnetic field strength values are taken between 0 G and 1000 G whilst inclination values are chosen in between 0$^{\circ}$ and 90$^{\circ}$ or 90$^{\circ}$ and 180$^{\circ}$ maintaining the polarity of the field as determined from the sign of the Stokes V profile. The result of the inversion is taken as the one with the best fit, that is, the minimum value of the $\chi^2$.

The inferred results using this strategy, and even if the signals are above the noise level, might not be reliable. The main reason for this is that given the field strengths involved, the spectral line used in this work could be in the so-called weak field regime. In this regime, the inversion is robust to the magnetic flux $\Phi=\alpha\,B\,\cos{\theta}$, where $\alpha$ is the filling factor, $B$ the magnetic field strength, and $\theta$ the magnetic field inclination, but not to the three parameters independently \citep{asensioramos2007}. One possible way to determine which signals allow a unique determination of the magnetic field inclination is to perform additional inversions for provided and fixed LOS inclinations and observe how the likelihood of the fit changes for these additional inversions. Here, we inverted all the positive polarity pixels again for a set of LOS inclinations (10$^{\circ}$, 20$^{\circ}$, 30$^{\circ}$, 40$^{\circ}$, 50$^{\circ}$, 60$^{\circ}$, 70$^{\circ}$, 80$^{\circ}$). In these inversions, we set the following parameters free to vary: the filling factor, the magnetic field strength, and the microturbulent velocity of the magnetic component. The remaining parameters (including the \mbox{non-magnetic} atmosphere) are taken from the result of the two-component inversion. If the $\chi^2$ value is similar for part of this set of LOS inclinations, that is, if there is more than one inclination value for which an equally good fit is achieved, this means that the inclination, filling factor, and magnetic field strength are degenerated and the only reliable quantity is the magnetic flux density. In other words, there are different combinations of these parameters that fit the profiles equally well. On the contrary, we consider that the LOS inclination is well determined. A sample of the pixels that fulfil this criterion for the disc centre case are highlighted in magenta in Fig. \ref{Fig:dcfovdata}.

Thus far, we have identified those pixels where the magnetic field inclination is well determined. However, even for those pixels for which the magnetic field inclination is reliable, the magnetic field strength and the filling factor (and also the microturbulent velocity) may still be coupled \citep[see e.g.][]{martinezgonzalez2006}. In order to verify the uniqueness of the inferred magnetic field strength we performed an additional test. We again performed several inversions for each pixel, this time fixing the magnetic field strength to 100, 300, 500, 700, 900, 1100, and 1300 G. The free parameters were now the microturbulent velocity of the magnetic component, the LOS magnetic field inclination, and the filling factor. As in the previous case, looking at the variation of $\chi^2$ for these inversions, we consider that whenever $\chi^2$ has a flat behaviour, the magnetic field strength is poorly determined, while if the $\chi^2$ does vary, the strength is considered to be well defined.

In this second test, we found that 93 \% of the pixels with a reliable determination of the inclination had a  poorly determined  magnetic field strength. Therefore, in the forthcoming analysis, we do not study the magnetic field strength, but we do analyse the magnetic field geometry and the magnetic flux density, which, according to this last test, is well determined within a 15 \% range around the average value (as determined using: $(\max{\phi}-\min{\phi})/|<\phi>|$).

\section{Results}
\label{Sec:results}

\subsection{Line-of-sight velocities}

\begin{figure*}
  \begin{center}
    \includegraphics[width=\textwidth]{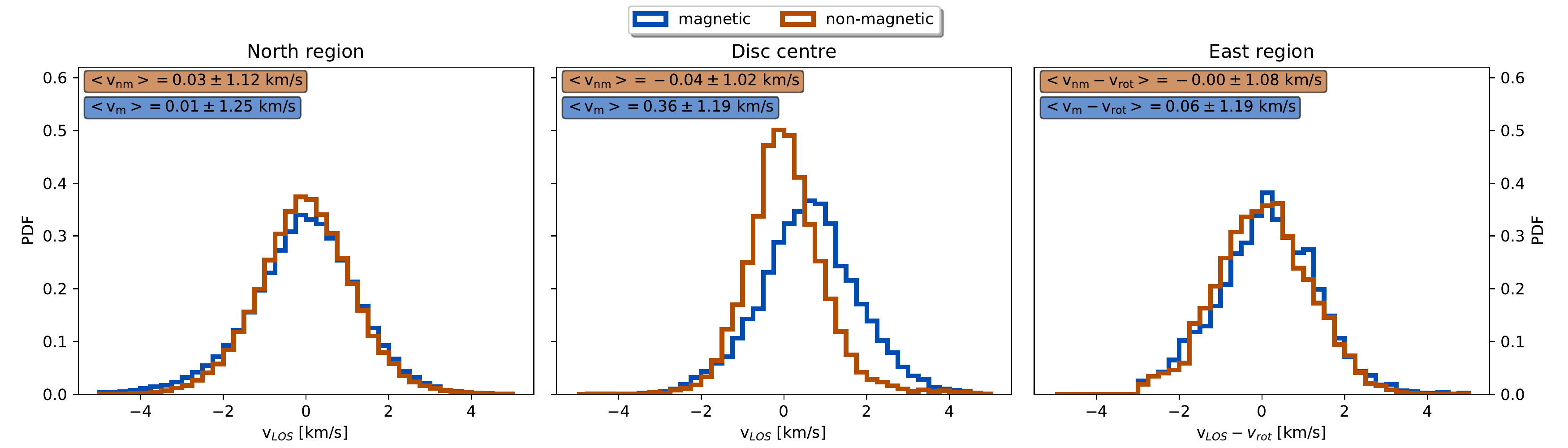}
    \caption{Line-of-sight  velocity PDFs inferred for each QS region. From left to right: Results for the north, disc centre, and east regions. The latter shows the actual  inferred LOS velocity minus the projected rotational velocity to the LOS at these longitudes. The LOS velocities for the magnetic component are shown in blue and for the non-magnetic component  in brown.}
    \label{Fig:sst_velocities}
  \end{center}
\end{figure*}

Before proceeding with the LOS velocity analysis, we subtract the velocity induced by the rotation of Earth. This effect comes from the fact that the Fabry-P\'erots calibration was performed in the morning (UT 07:10) while observations spanned until UT 10:30. To this aim, we followed \cite{plaskett1952},  calculating the variation with time of the Earth rotation velocity compared to the Sun at the latitude of the observatory ($\lambda=28^{\circ}\, 45^{\prime}\, 25^{\prime\prime}$) and subtracting it. The histograms of the corrected LOS velocities of both atmospheres are depicted in Fig. \ref{Fig:sst_velocities} for the three disc positions. 

The non-magnetic velocities at the disc centre (mid panel) are characterised by a symmetric distribution with a mean value of -0.04 km/s (upflow) and a standard deviation of 1.02 km/s. In contrast, the magnetic component velocity distribution is dominated by downflows, with a mean value of 0.36 km/s (downflow) and a standard deviation of 1.19 km/s, though the distribution shows larger tails. This offset of the magnetic component has already been found before \citep{grossmann-doerth1996, sigwarth1999, khomenko2003}. It is explained by the fact that magnetic patches tend to concentrate in intergranular lanes and in the borders of supergranular cells, in both cases dragged by convective motions at granular and supergranular scales. Therefore, because this analysis restricts to the pixels with polarimetric signals above three times the noise level, we might be biased to a set of pixels dominated by intergranular lanes (downflows). If this is the case for the magnetic velocity distribution, this argument should also be valid for the non-magnetic component, yet this is not seen. A possible explanation for this discrepancy is the different action of spatial smearing (i.e. the effects due to seeing, stray light, and binning) over the Stokes I profile (to which the non-magnetic atmosphere is most sensitive) and the Stokes Q, U, and V spectra (to which the magnetic atmosphere is most sensitive). In fact,  these smearing effects are expected to induce a mixing of information from the surroundings. For the Stokes I profiles (i.e. the non-magnetic atmosphere) this means that seeing effects and binning mixes intergranular profiles (downflows and redshifted Stokes I profiles) with surrounding granular light (upflows and blueshifted Stokes I profiles), potentially removing velocity information. The stray light effect also removes some velocity information, as for every pixel, it significantly contributes with a more or less constant intensity profile, provided by the extended wings of the optical PSF. In contrast, for Q, U, and V Stokes profiles (which mostly determine the magnetic atmosphere), because the strongest polarimetric signals are found in intergranular lines, the velocity information is not erased but its amplitude might be decreased as the surroundings might not have signals or are likely to have smaller amplitudes. This may also explain the narrower distribution found for the non-magnetic atmosphere as compared to the magnetic one, as the mixing of opposite velocities narrows the width of the distribution. Regarding the broad distribution found for the magnetic field component, this result is compatible with the studies mentioned above and the reason for its broad character is unclear, although it is likely that the presence of internal motions in magnetic structures and the influence of p-mode oscillations are playing a role.

At the east limb (right panel), the non-magnetic LOS velocity distribution has a mean value of -1.80 km/s and a standard deviation of 1.08 km/s. This large mean value is associated to the solar surface rotation velocity. For the average latitude ($<\lambda_{\odot}>=-0.13^{\circ}$) and longitude ($<l_{\odot}>=-71.18^{\circ}$) of the observed area, the projection to the LOS of the rotation velocity is -1.80 km/s, in close agreement with the average value found for the data. The velocity distribution of the magnetic component has a similar average (-1.74 km/s) and standard deviation (1.19 km/s). The fact that the average LOS velocity value at the limbs gets very close to zero is consistent with the average value found at disc centre taking into account projection effects. For a magnetic component characterised by predominantly vertical downflows, the projected component to the LOS would be smaller closer to the limb. The shape of the LOS velocity distribution close to the limb is broader than that of the disc centre. The reason for that could be the different effect of p-mode oscillations at this viewing angle as well as the presence of horizontal motions that contribute to the LOS velocities at this disc position. Very similar behaviour to that is seen at the north region (left panel). Both the magnetic and non-magnetic components are characterised by distributions centred at 0, in particular 0.01 km/s for the magnetic component and 0.03 km/s for the non-magnetic one, with a standard deviation of 1.25 km/s and 1.12 km/s, respectively. Line-of-sight velocities could therefore be explained by the same scenario for the three data sets.

\subsection{Line-of-sight magnetic flux}
\label{Sec:results_losflux}

\begin{figure*}
  \begin{center}
    \includegraphics[width=\textwidth]{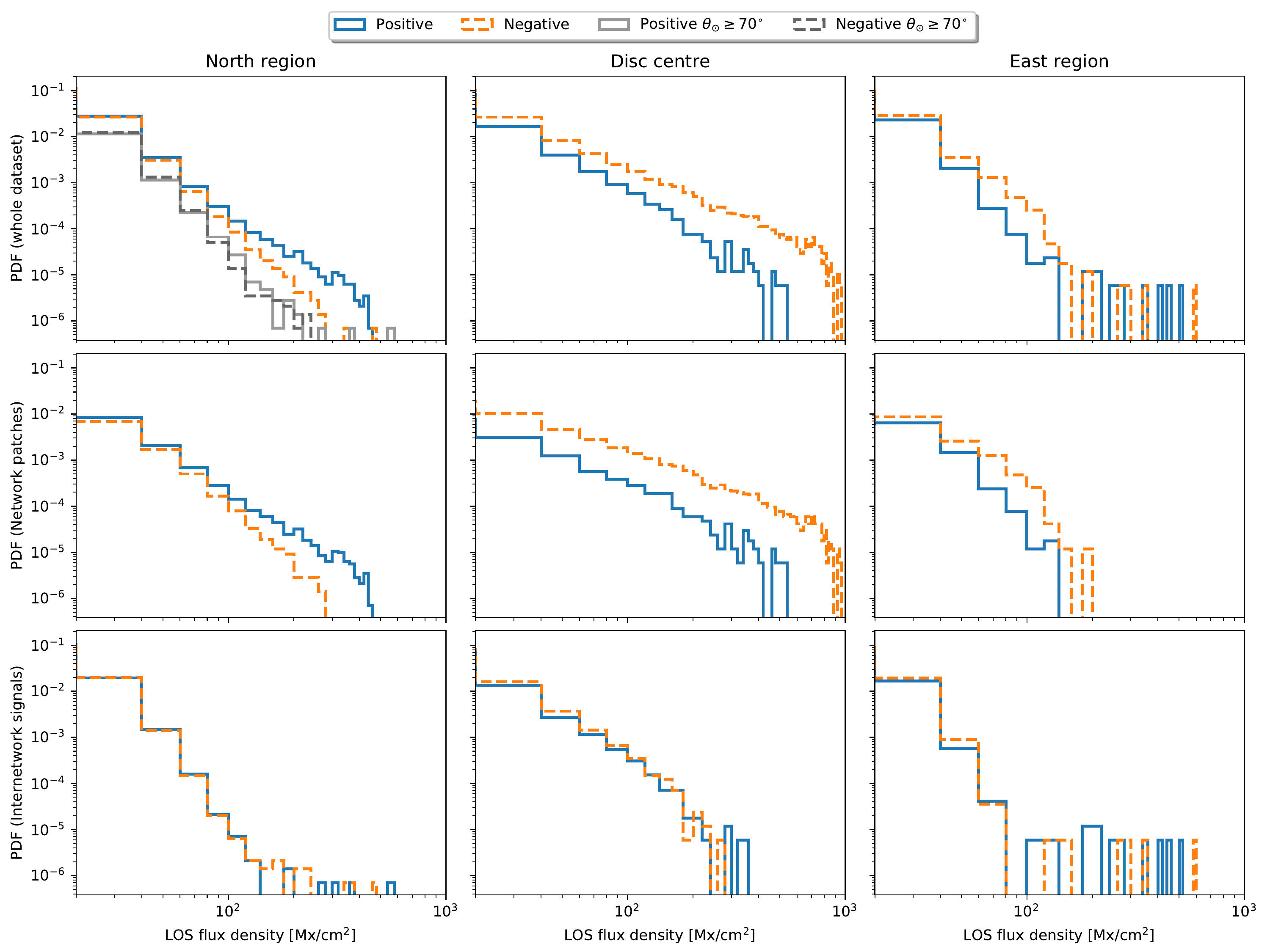}
    \caption{Line-of-sight magnetic flux inferred for the three QS regions. Top row, from left to right: LOS magnetic flux PDFs for north region, disc centre, and east-limb dataset. The positive tail is shown in solid blue, and the negative one is folded to positive values and shown by the orange dashed line. Middle panels follow the same representation for Network patches (see text for their definition) and bottom panels for the rest of the magnetic signals. The top-left panel additionally includes two grey lines to show the LOS magnetic flux above 70$^{\circ}$ heliocentric angle.}
    \label{Fig:sst_mflux}
  \end{center}
\end{figure*}

The PDFs of the LOS magnetic flux density ($\Phi$) for the three disc positions are shown in Fig. \ref{Fig:sst_mflux}. The three observed regions have a polarity imbalance (top row), which is positive for the north region, with a mean value of $0.36\pm0.02$ Mx/cm$^2$ and negative for disc centre and the east region with $-4.40\pm0.10$ Mx/cm$^2$ and $-1.28\pm0.06$ Mx/cm$^2$, respectively. Here we consider Network and Internetwork magnetic fields simultaneously, which are known to have different physical properties \citep[see e.g.][and references therein]{dewijn2009}. In order to isolate the contribution coming from each QS component, we identify as Network those areas with any (Q, U, or V) polarisation signals above 0.5 \% $I_{c}$. We do so because, depending on the viewing angle, strong vertical fields (specific to Network fields) give rise to different polarimetric properties: viewed from above, this kind of magnetic field has strong circular polarisation signals, while, from the side, they show strong linear polarisation signals. Additionally, we also include the surrounding polarimetric signals (two pixels)  in these `Network' patches, as they likely belong to the same feature, even when their polarimetric amplitude does not reach our threshold. The rest of the signals are assumed to belong to Internetwork areas. The middle and lower rows of Fig. \ref{Fig:sst_mflux} show the Network and Internetwork $\Phi$ distributions, respectively.

A common property to all the observed regions is that "Network" fields are in clear polarity imbalance, even when they represent a minor fraction of the magnetic signals (12.5 \%, 23.7 \% and 14.3 \% in the north, disc centre and east regions, respectively). In contrast, Internetwork fields are much closer to polarity balance. This latter result is in agreement with previous studies \citep{lites2002, khomenko2003, lites2011, martinezgonzalez2008}. In general (top row), the tails of the distributions do not reach  fluxes as strong as those at the disc centre. More specifically, at the east region, the distribution has a cutoff at lower magnetic fluxes than at the north region. This is due to the fact that the north pole observation covers a broader range of heliocentric angles than the map at the east limb. If we limit the north region to the heliocentric angles covered at the east region (top-left panel in grey lines), then the distribution more closely resembles that observed at the east region. This smaller LOS magnetic field flux density at limb regions might be related to the fact that, when observing close to the limbs, spectra are coming from higher up in the atmosphere, where the magnetic field is smaller. It is possible to make a simple calculation to gain some insight into this effect. For instance, for the east-limb dataset, $\mu=0.27$ and $\mu=0.37,$ we can assume that continuum is formed at optical depths with $\tau=\mu,$ and so close to the limb, the continuum we record comes from approximately 60-70 km higher in the atmosphere than the continuum from disc centre. This will also be the case for the spectral line as it is a weak line. If we further assume that the magnetic pressure has to be the same as the gas pressure, the magnetic field strength is proportional to $\exp{-z/{2\,H}}$, where $H$ is the height scale value, approximately 140 km. In such a simplified scenario,  the magnetic field strength for east data should be $\exp{-0.25}\approx0.78$ of the one measured for disc centre, which is consistent with the measured values.

Interestingly, the mean LOS magnetic flux of the polar region (positive) matches that of the building up polarity of cycle 24. For instance, in Fig. 2 of \cite{sun2015} one can see that, before the polarity reversal in 2013, the north polar region had an average negative polarity and, after 2013, it had a very weak but positive one. Another very interesting fact is that, if we calculate the average of the LOS magnetic flux value for solar latitudes above $70^{\circ}$, then $<\Phi>=-0.39\pm0.03$ Mx/cm$^2$ (where the uncertainty refers to that of the average estimation for 170367 cases), which means that the highest latitudes still have the old (opposite) polarity. This finding is consistent with the global picture where polar regions change their polarity with incoming opposite polarity from low/mid latitudes. Therefore, for these observations, which were made close to the polar region polarity reversal of mid-2013 \citep[see e.g.][]{pastoryabar2015}, a partial polarity change is reasonable.

\subsection{Line-of-sight magnetic topology}

Hereinafter, we restrict the analysis to those pixels where the LOS magnetic field inclination is well determined. As seen in Sect. \ref{Sec:inversion}, in most cases, we still cannot determine a reliable value for the magnetic field strength, so this parameter is avoided in this section.

\begin{figure*}
  \begin{center}
    \includegraphics[width=\textwidth]{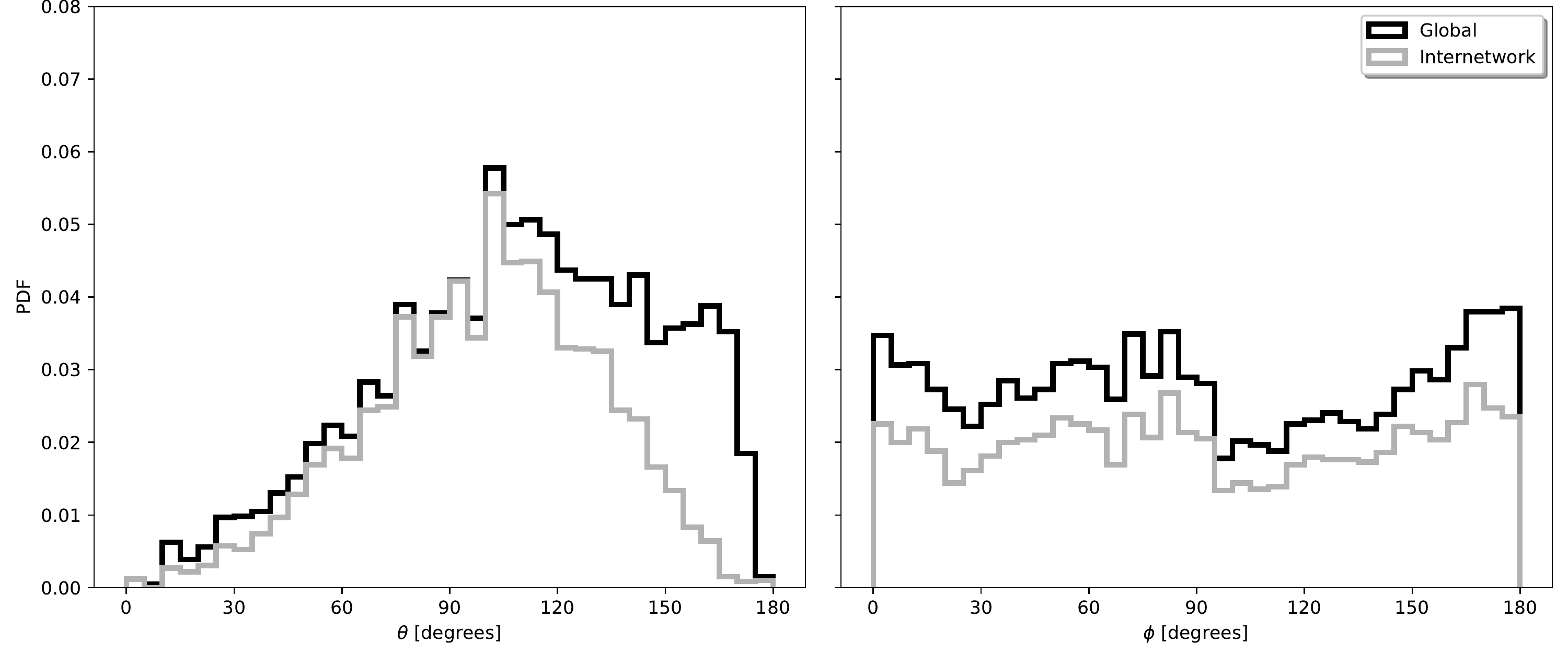}
    \caption{Disc-centre magnetic field orientation. From left to right: PDFs of the LOS magnetic field inclination and azimuth.}
    \label{Fig:sst_magnetics_los_dc}
  \end{center}
\end{figure*}

Figure \ref{Fig:sst_magnetics_los_dc} displays the PDFs for the LOS magnetic field inclination ($\theta$, hereafter) and azimuth ($\phi$), between $0^{\circ}$ and $180^{\circ}$, at the disc centre. At this disc position, interpretation of the magnetic field topology is more straightforward as the LOS coincides with the solar vertical.

\begin{figure*}
  \begin{center}
    \includegraphics[width=0.99\textwidth]{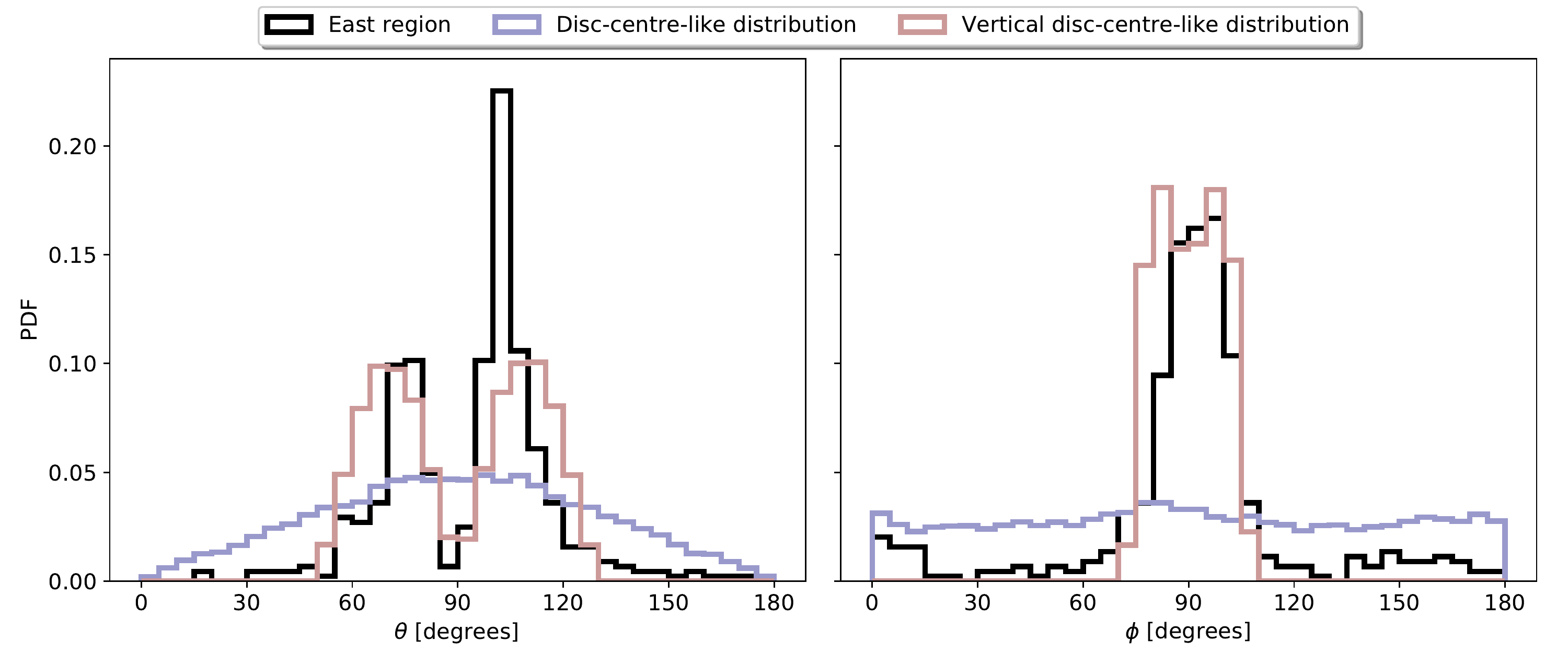}
    \includegraphics[width=\textwidth]{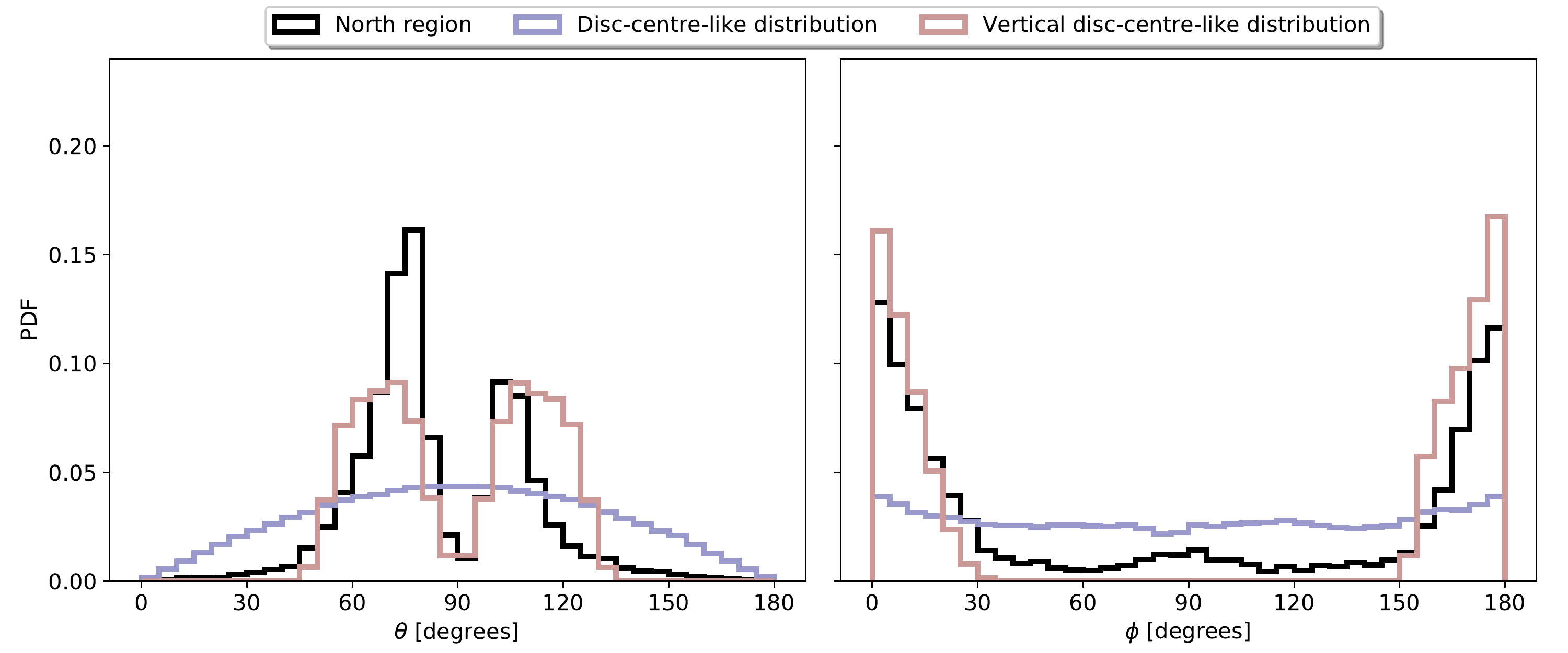}
    \caption{Limb datasets LOS magnetic field orientation. In solid black and from left to right: PDFs of the LOS magnetic field inclination and azimuth of the east region data set (top row) and north one (bottom row). The observed disc centre magnetic field topology is shown in blue, as it would be seen at limb regions when taking into account geometrical effects. A subset (vertical fields) of the observed disc centre magnetic field topology is shown in red.}
    \label{Fig:sst_magnetics_los_both}
  \end{center}
\end{figure*}

 At disc centre, $\theta$ has a polarity imbalance with a dominant negative polarity. This was already seen for the LOS magnetic field flux density distribution (Fig. \ref{Fig:sst_mflux}) and it is associated to the presence of strong magnetic patches of this polarity belonging to the Network. These strong fields have inclinations close to the LOS ($\sim180^{\circ}$), and therefore perpendicular to the surface. As mentioned above, the physical properties of Network and Internetwork fields are known to be different, and therefore we display the $\theta$ and $\phi$ distributions belonging to Internetwork elements in grey in the same figure. Internetwork $\theta$ shows a more balanced polarity distribution (even though negative polarities are more abundant) with a preference for horizontal fields ($\theta\sim90^{\circ}$). Regarding the orientation over the solar surface (right panel of Fig. \ref{Fig:sst_mflux}), we found that both components depict roughly equally probable orientations over the solar surface.

The results shown here for the Network component are consistent with previous results \citep[see e.g. Sect. 2.2 in][and references therein]{borrero2017}. Regarding the Internetwork, the orientation of these fields remains unclear, with some works \citep{martinezgonzalez2008b, asensioramos2009, stenflo2010, faurobert2015} pointing to more isotropic distributions and others \citep{orozcosuarez2007, orozcosuarez2007b, lites2007, lites2008, borrero2013} retrieving more horizontal distributions. Our results seems to favour the latter, yet they must be taken with caution as, by selection, we might be biased to this type of distribution, owing to the fact that we are analysing the magnetic fields that have a well defined orientation. At this disc position and for this observational setup, these selection effects happen, in general, under two scenarios: First, significantly strong linear polarisation profiles are detected. These signals induce a strong preference for horizontal orientations, as we infer here. This does not mean that there are not other magnetic field orientations but that we can uniquely determine the orientation for signals with sufficiently strong linear polarisation signals (horizontal fields). Second, very strong fields exists where the magnetic splitting is well above the Doppler width of the spectral line. This occurs, for this spectral region and line, for magnetic field strengths above $\sim1200$ G, which, in the QS region, are due to vertical fields, i.e. the Network fields \citep{borrero2017} -- this case does not appear in the Internetwork fields.

The $\theta$ and $\phi$ for the east (top row) and north regions (bottom row) are presented in Fig. \ref{Fig:sst_magnetics_los_both}. In both cases, the observed FOV is far from disc centre and so the LOS reference frame is not the local reference frame (LRF). That is why we need to rotate the reference frame from LOS to LRF, which in principle involves two rotations of known angles. However, there exists a spectropolarimetric intrinsic ambiguity that complicates this step. This is known as the 180$^{\circ}$ LOS magnetic field azimuth ambiguity, and implies that each pixel has two possible magnetic field configurations depending on the value chosen for the LOS azimuth.

Before proceeding with the disambiguation, some information about the degree of compatibility between the observed distributions at the disc centre and at the limbs might be retrieved. To this aim, we follow the same strategy as in \cite{pastoryabar2018}. In this method, we use the disc-centre inferred topology as our reference for any QS region, in particular, for that of the observed east (north) area. Knowing the LRF topology (the one inferred at disc centre) and the coordinates of every pixel in the limb dataset, we can build the equivalent LOS topology as it would be observed for a disc-centre-like distribution. This step only takes into account the change in the viewing angle, while it neglects radiative transfer effects or signal mixing as the area covered by our pixel size due to the limb foreshortening is larger; nevertheless, it provides some meaningful insight into the inferred magnetism.

The result of this test is shown in blue in Fig. \ref{Fig:sst_magnetics_los_both}. It is evident that the observed distributions at any limb region (black) do not agree with the observed ones at disc centre as seen at these disc positions (blue). In contrast to the smooth LOS $\theta$ expected from disc-centre data, the ones at the limbs show two clear peaks at around 75$^{\circ}$ and $105^{\circ}$. These LOS inclination values are close to the heliocentric angle of the observed areas, which are values expected for magnetic fields close to the local vertical. In order to further consider this possibility, we repeat the same compatibility test as before but now for a subset of the fields inferred at disc centre, that is, those with inclinations below $15^{\circ}$ or above $165^{\circ}$. The result of this new test is presented in red in Fig. \ref{Fig:sst_magnetics_los_both}. Now the agreement between the observed distributions at the limbs and the ones one would expect to see at those regions if the LRF magnetic fields were vertical is much closer. Furthermore, this second test also shows that the observed LOS $\phi$ distribution has a very clear orientation in specific directions. As the SST calibration procedure sets positive Q in the solar north--south direction and we set the 0$^{\circ}$ LOS azimuth along this direction, at the east, the azimuths take values close to 90$^{\circ}$, while at the north region, they are at 0$^{\circ}$ and 180$^{\circ}$, i.e. they are aligned along the radial direction.

\subsection{Local-reference-frame magnetic topology}

\begin{figure*}
  \centering
    \includegraphics[width=1.\textwidth]{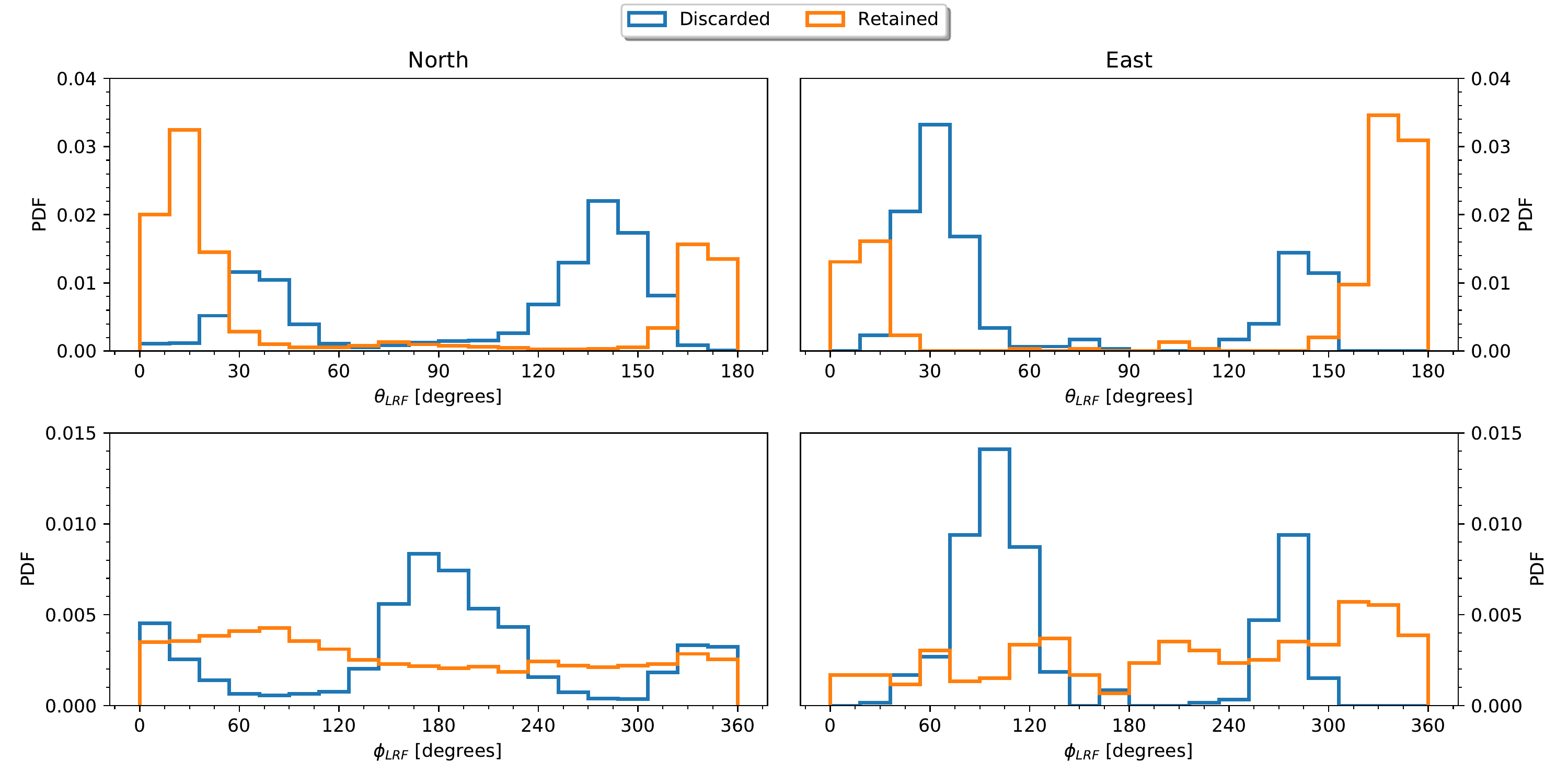}
    \caption{Limb datasets LRF magnetic field orientation. Top row: PDF of the LRF magnetic field inclination. Bottom row: PDF of the LRF azimuth. From left to right: Results for north and east regions. The distributions to be retained after the LOS to LRF rotation problem has been solved are shown in orange, and the discarded solution is shown in blue.}
    \label{Fig:sst_lrf_distribution_both}
\end{figure*}

For a proper comparison between the limb datasets and those at disc centre it is convenient to rotate from the LOS to the LRF. To do so, we follow the same method detailed in \cite{pastoryabar2018}. To that aim, we identify individual magnetic structures where the LOS magnetic field azimuth and inclination are reliably inferred (292 structures for the north region and 23 for the east one). Subsequently and under the assumption that the magnetic field vector in the LRF is smooth all along the structure, the 180$^{\circ}$ LOS azimuth ambiguity per pixel translates into two possible magnetic field topologies for each structure in the LRF. These two solutions for each magnetic structure in the LRF are physically reasonable. In order to pick one of the two possible morphologies for each structure, we demand that the final LRF azimuth ($\phi_{LRF}$) distribution, given by the whole set of structures identified, at each disc position, be as flat as possible. This procedure is done iteratively switching between the two possible solutions for the LRF magnetic field orientations retrieved for each structure and checking for the flatness of the azimuth distribution of the whole dataset as a guide. This criterion is based on the $\phi_{LRF}$ distribution inferred at the disc centre, and also as we found no argument to have a preferred magnetic field orientation for a set of QS magnetic field structures over the solar surface. This way, the retrieved topology (``Retained'') and the discarded one (``Discarded'') for both limb datasets are shown in Fig. \ref{Fig:sst_lrf_distribution_both}.

Both east and north QS regions are compatible with a roughly flat LRF azimuth distribution. For this orientation over the solar surface, the LRF inclination distributions are dominated by vertical fields of mixed polarities. The discarded solution, although physically possible, is given by a very specific orientation,  and corresponds to mid-inclined fields (around 40$^{\circ}$ from the local vertical). In addition, these fields are aligned over the solar surface in the direction that connects the position of the structure with the disc centre as seen by the observer, i.e. 0$^{\circ}$, 180$^{\circ}$, and 360$^{\circ}$ for the north region and 90$^{\circ}$ and 270$^{\circ}$ for the east one. Such a magnetic field configuration is a very singular topology, which depends on the disc position and is aligned with the radial direction as seen by the observer. This fact makes this solution very unlikely.

From this analysis, both limb datasets are similar to each other and their magnetism is compatible with vertical magnetic fields. This result is partially in agreement with previous studies \citep{blancorodriguez2007, ito2010, blancorodriguez2010, jin2011, shiota2012, pastoryabar2018} but here we are missing the second (weak and with no clear magnetic field inclination) component reported by these authors. We highlight the fact that in this work we detect many more magnetic fields (polarimetric signals above $3\sigma$) than the ones analysed in this section; see for instance the large strong circular polarity structure at $x\sim3^{\prime\prime}$, $y\sim12^{\prime\prime}$ in Fig. \ref{Fig:elfovdata}. Here, even though the circular polarisation signals are present and clear as a coherent structure, the lack of linear polarisation signals (notice that one cannot see this structure at all in the linear polarisation map) and the absence of sufficiently strong magnetic fields prevents us from deciphering the orientation or the strength of these magnetic fields as we have reported in Sect. \ref{Sec:inversion}, and so we cannot further discuss their physical properties.

\subsection{Local-reference-frame magnetic flux}

Now that we have rotated the inferred parameters from the LOS to the LRF, we can calculate the LRF magnetic flux harboured by each structure (292 magnetic patches at the north polar region for which the rotation from LOS to LRF is performed). This parameter is the result of the 
previously introduced magnetic flux density ($\Phi=\alpha\,B\,\cos{\theta}$) multiplied by the pixel area and integrated over the whole patch. In order to calculate the pixel area, the  foreshortening effect acting on  limb observations  must be taken into account. We reiterate the fact that even when the magnetic field strength is not well determined (see Sect. \ref{Sec:inversion}), we see that the magnetic flux is a well constrained parameter (up to a 15 \%).

\begin{figure*}
  \begin{center}
    \includegraphics[width=\textwidth]{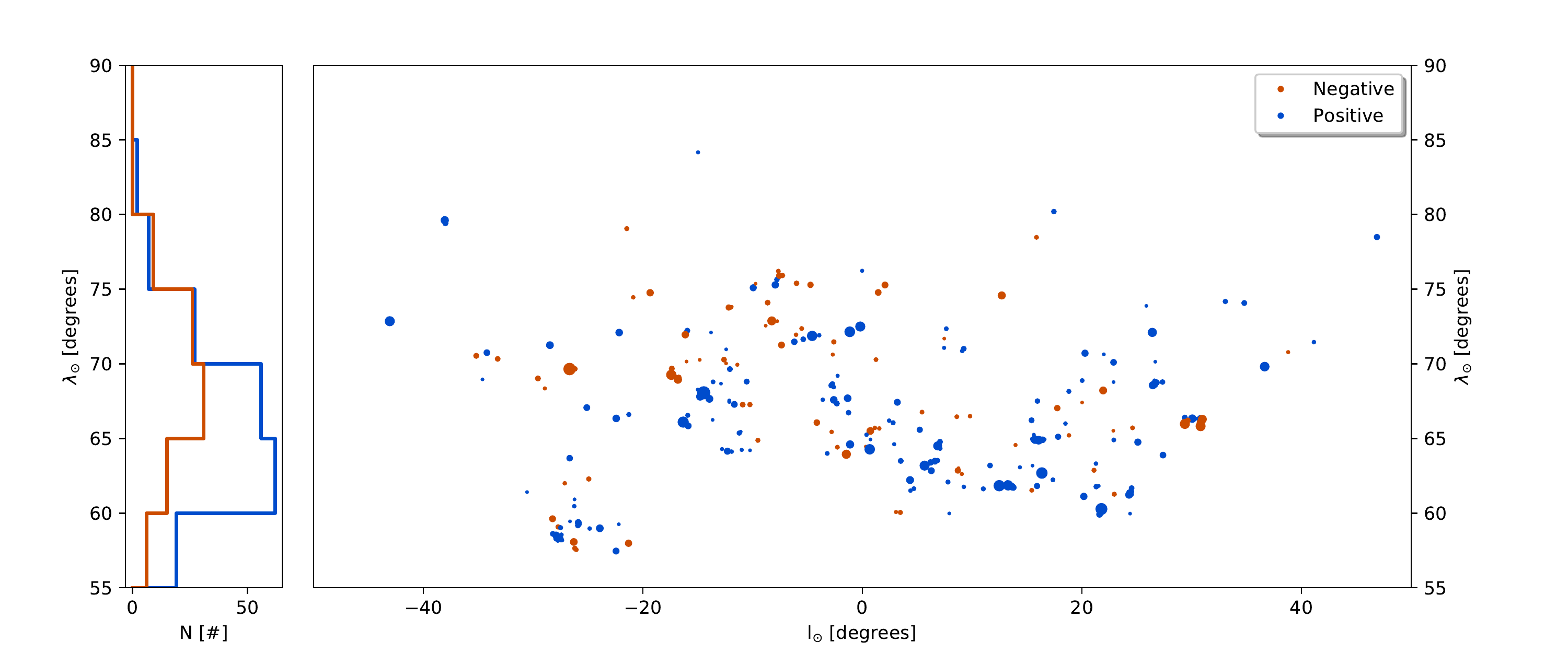}
    \caption{Spatial distribution of the magnetic patches observed in the north region. Right panel: Solar heliospheric longitude and latitude distribution of the local reference magnetic flux for the magnetic patches found in the north region data set. Blue colour stands for positive (outgoing from the Sun) magnetic fields. Negative (ingoing to the Sun) magnetic fields are plotted in brown. The size of the dots is proportional to the decimal logarithm of the LRF flux of each patch. Left panel: Histogram of the number of magnetic patches as a function of the solar latitude according to their LRF magnetic polarity. The colour code is preserved for the sake of clarity.}
    \label{Fig:sst_lrf_polarity_spatial_distribution}
  \end{center}
\end{figure*}

In Fig. \ref{Fig:sst_lrf_polarity_spatial_distribution}, we explore the polarity distribution of the magnetic patches rotated to the LRF over the covered north polar area. Positive patches dominate at lower latitudes (below latitude $70^{\circ}$). This result is clearer in the histogram on the left, where the number of positive and negative LRF magnetic patches with latitude are depicted. Above this latitude threshold, the number of magnetic patches with positive and negative polarity is relatively balanced.

\begin{figure*}
  \begin{center}
    \includegraphics[width=\textwidth]{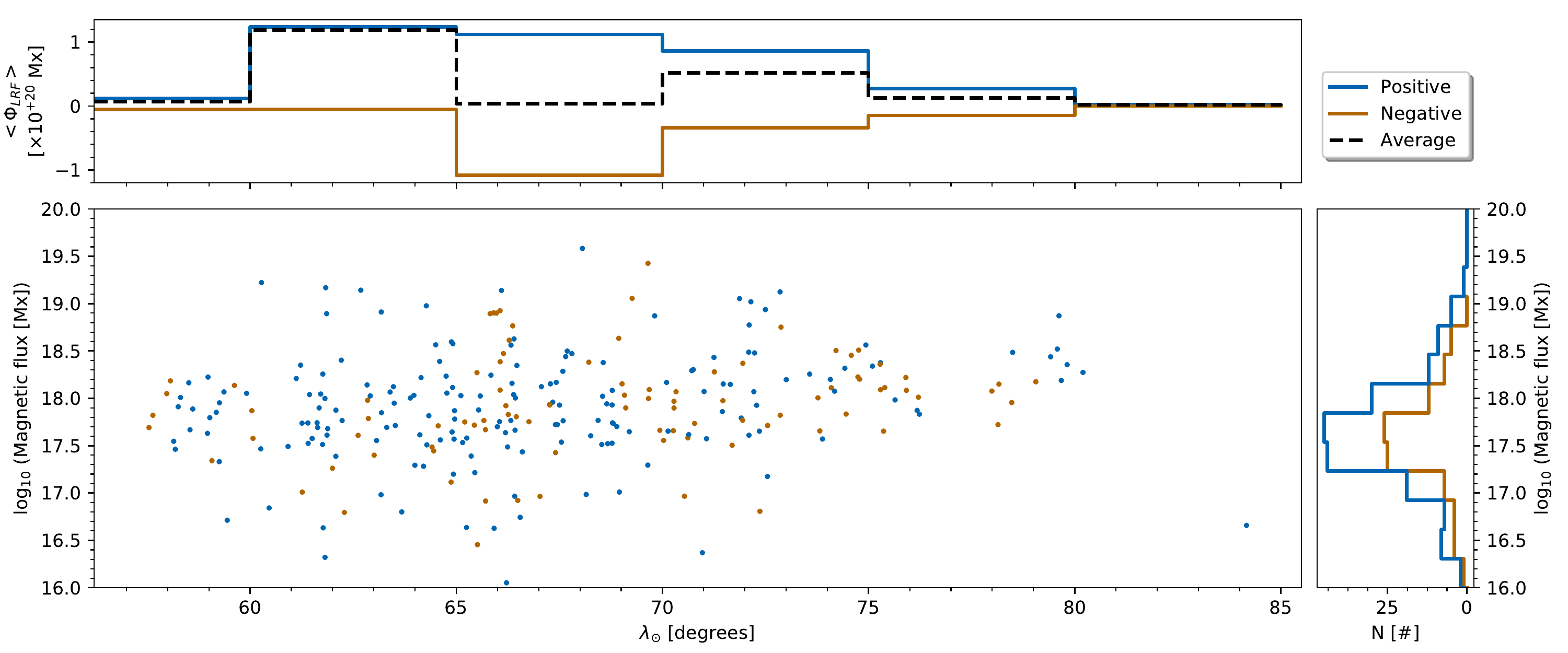}
    \caption{Local-reference-frame magnetic flux distribution depending on the heliospheric latitude (bottom-left panel). Blue colour stands for positive flux (outgoing the solar surface) and brown for negatives values. Right lower panel: LRF histograms. Top panel: Averaged positive, negative, and total LRF flux ($\Phi_{LRF}$) as a function of the solar latitude. The colour code is preserved for the sake of clarity.}
    \label{Fig:sst_lrf_polarity_distribution}
  \end{center}
\end{figure*}

Regarding the distribution with latitude of these magnetic patches depending on the LRF magnetic flux that they harbour, we find that most of the studied structures carry an LRF magnetic flux of between $10^{17}$ Mx and $10^{19}$ Mx (see Fig. \ref{Fig:sst_lrf_polarity_distribution}). As we move from low latitudes to upper ones, we find that the smallest (as compared to the observed average value) LRF magnetic fluxes ($\Phi_{LRF}<10^{17}$ Mx) and the largest ($\Phi_{LRF}>10^{19}$ Mx) ones progressively disappear. This way, above a latitude of 73$^{\circ}$  these magnetic fluxes are absent (except for a structure around 84$^{\circ}$ latitude).

One may expect a loss of sensitivity to the weakest fluxes of the distribution when observing closer to the limb as foreshortening effects become stronger, yet this is not the case for the strongest magnetic structures. This might be indicative that these large magnetic flux structures are absent from the most polar region. On the right of the figure, we show the number of structures of each polarity for the various LRF magnetic fluxes. The whole distribution is dominated by positive polarity structures. The sharp fall in the number of structures (of both polarities) around $10^{17}$ Mx might be indicative of our lower limit sensitivity, as it is observed that the smaller the magnetic flux, the larger the number of structures \citep[][down to $10^{16}$ Mx]{kaithakkal2013}.

If we focus on the average LRF magnetic flux over the observed area we find that it is positive both for the whole polar area observed and for each latitude bin considered (see top panel in Fig. \ref{Fig:sst_lrf_polarity_distribution}). This is in clear contrast with the results obtained for the LOS magnetic flux density (see Sect. \ref{Sec:results_losflux}), which was negative for the polar area above $70^{\circ}$ latitude. This difference comes from the fact that the magnetic fields considered here, that is the ones for which we can infer the LOS inclination accurately enough to perform the rotation from LOS to LRF, are much less than the ones considered when studying the LOS magnetic flux density (all the magnetic signals). This way, depending on the proxy we use to estimate the polarity reversal of the polar region, we may or may not conclude that the closest area to the solar pole has already changed its polarity, at least in between $70^{\circ}$ and $80^{\circ}$. This is because the average LOS flux density has the sign of the old dominant polarity $\Phi_{LOS}\in(\lambda_{\odot}=70^{\circ},\lambda_{\odot}=80^{\circ})<0$ whilst the LRF flux of the strongest magnetic structures of that region, which are known to determine the polar region dominant polarity \citep{shiota2012}, has a positive average value ($\Phi_{LRF}\in(\lambda_{\odot}=70^{\circ},\lambda_{\odot}=80^{\circ})>0$). This result emphasises the importance of the interpretation of the LOS magnetograms, at least close to maximum activity. Additionally, (most) coronal magnetic field extrapolations rely on the photospheric magnetism and so this point is of critical importance because depending on the proxy used to estimate the photospheric magnetism, the result for upper layers might change completely.

\section{Discussion and conclusions}

We analysed the magnetism at the north polar cap close in time to a polarity reversal using full spectropolarimetric data for a photospheric spectral line. We compared the physical parameters inferred at the polar area and in two QS areas at low latitudes: at disc centre and at an equatorial limb. This strategy allows the comparison of polar magnetism and low-latitude QS, minimising the role of projection effects over the observations and the posterior inferred parameters.

We find that neither the LOS velocity nor the LOS magnetic flux, when studied for all the polarimetric signals detected, show any particular behaviour at the polar region. For the velocity, we find that both limb cases are compatible with the disc centre LOS velocity for the magnetic field component, which has an average downflow of $0.36$ km/s. Also, in all the observed FOVs, the average LOS magnetic field flux is determined by the Network component, which represents a minor fraction (around 10-25 \%) of the polarimetric signals analysed. The rest of the polarimetric signals are in LOS polarity balance.

For the polarimetric signals for which we find a unique magnetic field geometry, we also find that both polar regions are populated by vertical magnetic fields. This is partly consistent with what we found for the disc centre. There we find that magnetic fields are either vertical or horizontal. However, these results must be considered with caution because those pixels for which we are able to infer the magnetic field geometry in a unique manner are the pixels for which we detect sufficiently strong linear polarisation signals or the Zeeman splitting is larger than the Doppler splitting (or both). Only under these two conditions were we able to infer a unique geometry for the magnetic field. At disc centre, this means that our criteria might be fulfilled by horizontal fields that give linear signals and to strong fields which at this disc position are vertical. In contrast, at the limbs, the viewing angle changes and we find that strong fields are the only ones that give rise to linear polarisation signals, while no other signals allow a unique inference and so we get different geometries between disc centre and limb datasets. Nevertheless, it is important to point out that both limb regions behave in the same way, namely, there is no distinctive feature at the polar region; and there exists a common scenario that can explain the limb datasets and part of the observed magnetism at disc centre. The observation of vertical fields is in agreement with previous full spectropolarimetric analyses; \cite{ito2010, blancorodriguez2010, jin2011, shiota2012, quinteronoda2016} and \cite{pastoryabar2018}. In some of the previous studies, additional components were found. \cite{ito2010, jin2011} and \cite{shiota2012} found an additional magnetic field component of horizontal fields, \cite{blancorodriguez2010} found that in addition to vertical fields there exists an isotropic component, and \cite{pastoryabar2018} found an additional component compatible with the presence of spatially unresolved small-scale magnetic loops. In the present study, apart from the fields for which we have been able to determine their geometry, there are many more magnetic signals that might be related with the ones found in these latter studies, yet our data lack accurate enough polarimetric sensitivity to further address this point.

Focusing on the spatial and magnetic properties of the magnetic patches over the polar region for which the rotation from LOS to LRF is possible, we find that these structures are characterised by large flux concentrations (with $\Phi_{LRF}\geq 10^{16}$ Mx). These magnetic patches show, for all the range of $\Phi_{LRF}$ detected, a polarity imbalance with dominant positive polarity. As the largest magnetic field concentrations define the dominant polarity of the polar region \citep{shiota2012}, this might suggest that the north polar region has already built up its new dominant polarity. Also, we find that these magnetic patches are mostly seen in the range of $\lambda_{\odot}\in(60^{\circ},73^{\circ})$ with only a few of them above this upper limit. This could be related to the fact that observations took place close to a maximum of activity, i.e. close to the period in which polar caps reverse their polarity. In the standard polar polarity reversal model, during and after the polarity reversal of the polar region, the new building flux arrives at the polar regions from lower latitudes \citep{benevolenskaya2004}, and so, our results might be explained by a scenario where these strong magnetic flux concentrations have not  yet reached the uppermost latitudes. Finally, a remaining open question concerns the fact that we find opposite dominant polarity signs for the average LOS flux density above $\lambda_{\odot}=70^{\circ}$ and the LRF magnetic flux. This point is important because coronal magnetic field extrapolations strongly rely on the magnetism measured in the photosphere of the polar regions \citep[][]{petrie2015}, which is sometimes estimated from LOS magnetograms together with some assumption on the topology of the magnetic field. In this context, the Solar Orbiter mission, and more specifically the Polarimetric and Helioseismic Imager \citep[SO/PHI,][]{solanki2019} onboard, is expected to provide invaluable insight. The SO/PHI will record full spectropolarimetric data (allowing the inference of the photospheric magnetic field vector) for the polar regions from a privileged viewing point, that is, from outside the ecliptic (up to 24$^{\circ}$during the nominal phase and 33$^{\circ}$ in the extended one), mitigating the problems inherent to polar region observations from Earth or its surroundings.

\begin{acknowledgements}
APY would like to thank Jaime de la Cruz Rodriguez and Pit S\"utterlin for their assistance setting up and running SST/CRISP observations. We thank the anonymous referee for their comments that significantly improved the manuscript. Financial support by the Spanish Ministry of Economy and Competitiveness and the European FEDER Fund through projects AYA2014-60476-P (Solar magnetometry in the era of large solar telescopes) and AYA2014-60833-P (Spectropolarimetry: a window to stellar magnetism) is gratefully acknowledged. APY is supported by the German Government through {\it Deut\-sche For\-schungs\-ge\-mein\-schaft, DFG\/} project number 321818926 ``STOK3D Three dimensional Stokes Inversion with magneto-hydrostationary constraints". The Swedish 1-m Solar Telescope is operated on the island of {\it La Palma} by the Institute for Solar Physics of Stockholm University in the Spanish {\it Observatorio del Roque de los Muchachos} of the {\it Instituto de Astrof\'{\i}sica de Canarias, IAC}. The Institute for Solar Physics is supported by a grant for research infrastructures of national importance from the Swedish Research Council (registration number 2017-00625). This paper made use of the IAC Supercomputing facility HTCondor (\url{http://research.cs.wisc.edu/htcondor/}). NSO/Kitt Peak FTS data used here were produced by NSF/NOAO. 
\end{acknowledgements}

%
   \bibliographystyle{aa} 
   \bibliography{biblio} 
%

\end{document}